\documentstyle[amssymb,aps,prb,floats,twocolumn,epsfig]{revtex}

\begin{document}

\wideabs{\title{Vertical transport and electroluminescence in
InAs/GaSb/InAs structures:
\\GaSb thickness and hydrostatic pressure studies}

\author{M. Roberts, Y.C. Chung, S. Lyapin, N.J. Mason, R.J. Nicholas, P.C. Klipstein} \address{Department of Physics,
Oxford University, \\Clarendon Laboratory, Parks Rd., Oxford, OX1
3PU, U.K.}

\date{\today}
\maketitle

\begin{abstract}
We have measured the current-voltage (I-V) of type II
InAs/GaSb/InAs double heterojunctions (DHETs) with 'GaAs like'
interface bonding and GaSb thickness between 0-1200\,\AA. A
negative differential resistance (NDR) is observed for all DHETs
with GaSb thickness $>$ 60\,\AA\, below which a dramatic change in
the shape of the I-V and a marked hysteresis is observed. The
temperature dependence of the I-V is found to be very strong below
this critical GaSb thickness. The I-V characteristics of selected
DHETs are also presented under hydrostatic pressures up to 11
kbar. Finally, a mid infra-red electroluminescence is observed at
1 bar with a threshold at the NDR valley bias. The band profile
calculations presented in the analysis are markedly different to
those given in the literature, and arise due to the positive
charge that it is argued will build up in the GaSb layer under
bias. We conclude that the dominant conduction mechanism in DHETs
is most likely to arise out of an inelastic electron-heavy-hole
interaction similar to that observed in single heterojunctions
(SHETs) with 'GaAs like' interface bonding, and not out of
resonant electron-light-hole tunnelling as proposed by Yu et al. A
Zener tunnelling mechanism is shown to contribute to the
background current beyond NDR.
\end{abstract}

}

\section{Introduction}
\begin{figure}[h]\centering \epsfxsize=85mm
\epsffile{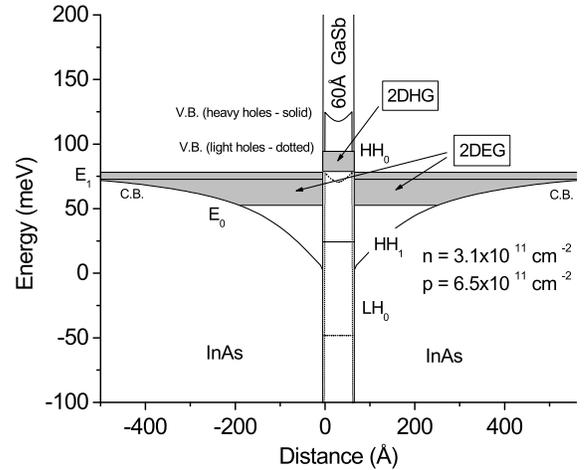} \vspace{0.2cm} \caption{Equilibrium
band profile of 60\,\AA\, GaSb DHET } \label{fig:Profile 60A}
\end{figure}

There has been considerable interest recently in resonant
tunnelling devices involving coupling between the conduction and
valence bands of different materials across heterointerfaces. Much
of this study has focussed on InAs/GaSb/AlSb structures owing to
the 'broken gap' band alignment at the InAs/GaSb interface.
Devices exploiting the strong interband coupling shown in these
structures include resonant tunnelling diodes \cite{Soderstrom
1989a} and interband cascade lasers \cite{Lin 1997}. Despite rapid
progress and extensive modelling, the details of the conduction
mechanisms in even the simplest structures still remain poorly
understood. Following a detailed study of the conduction in
InAs/GaSb SHETs (single heterojunctions) \cite{Khan-Cheema 1996},
where hydrostatic pressure was used to vary the band overlap and
hence the interband coupling, this work investigates the case of
the InAs/GaSb/InAs DHET (double heterojunction), shown by the band
profile in Fig. \ref{fig:Profile 60A}. Both SHETs and DHETs show a
region of negative differential resistance in their IV
characteristic and abnormally high background currents of
unconfirmed origin beyond NDR.

NDR arising out of interband coupling in single barrier structures
was proposed \cite{Chow 1986} and demonstrated \cite{Chow 1987} by
Chow et al. for the staggered lineup HgCdTe/CdTe/HgCdTe DHET, and
subsequently by Beresford et al. \cite{Beresford  1989} and
Munekata et al. \cite{Munekata 1989} and S$\tt{\ddot
o}$derstr$\tt{\ddot o}$m et al. \cite{Soderstrom 1989b} from
InAs/AlGaSb/InAs DHETs. The NDR in this case can be explained by
consideration of the imaginary wave vector of the tunnelling
electrons in the barrier region using a two band
electron-light-hole k.p model \cite{Heremans 1986}. This shows
that the tunnelling initially becomes more difficult with
increasing bias as electrons strike the barrier with energies
further from the barrier valence band edge, thus resulting in a
region of NDR. NDR was subsequently observed in a broken gap
InAs/GaSb/InAs DHET by Taira et al. \cite{Taira 1989}, who
suggested that the current transport mechanism was ohmic
conduction from one InAs electrode to the other through the GaSb
valence band, which is cutoff at higher bias by GaSb band gap
blocking. A similar characteristic was observed by Luo et al.
\cite{Luo 1990} a year later and interpreted in terms of
conduction through transmission resonances arising out of the
mixing between InAs conduction band and GaSb valence band states.
Further investigation into the GaSb thickness dependence of the
I-V by Yu et al. \cite{Yu 1990} showed that the NDR is lost as the
GaSb thickness is reduced below 60\,\AA. According to their
calculations this loss in NDR coincides with the loss in the
electron-light-hole overlap at flat band. This led the authors to
conclude that the NDR arises out of a resonant electron-light-hole
tunnel current, and that the heavy-holes contribute only weakly to
the conduction. More detailed theoretical models have since been
presented in support of this conclusion, where electron-heavy-hole
coupling considerations are shown to only give rise to secondary
features in the theoretical IV characteristic \cite{Ting 1992}.

However, there are reports in the literature where the results are
difficult to interpret using the resonant electron-light-hole
tunnelling model. Takamasu et al. \cite{Takamasu 1994} have
applied magnetic fields parallel to the interfaces of
InAs/GaSb/InAs DHETs showing NDR, where the peak NDR bias gives
the opposite behaviour to that predicted. A detailed study of the
conduction mechanisms in the related InAs/GaSb SHET structures by
Khan Cheema et al. \cite{Khan-Cheema 1996} has shown that the
interface bonding ('InSb-like' or 'GaAs-like') plays a very
important role in the conduction, and that the NDR in the 'GaAs'
case is more likely to arise out of an inelastic
electron-heavy-hole interaction. In addition the origins of the
large background currents observed in both these structures and in
similar structures with AlSb barriers are not well understood. The
temperature \cite{Shen 1995} \cite{Chen 1990} and barrier
composition dependence \cite{Wagner 1995} of the valley current
suggest the possibility of light-hole tunnelling into the emitter
valence band at biases above 400mV, although this has not yet been
observed directly.

In all cases, the modelling is strongly dependent on the value
taken for the band overlap, which is believed to lie in the range
125-155meV and is dependent on the interface bonding \cite{Daly
1995a}. None of the reports in the literature include the effects
of strain on the heavy and light-hole valence band edges, which we
believe to be significant. The band profile models are also very
sensitive to the treatment of charge buildup in the
self-consistent potentials on either side of the type II
interface. Simple models assume a 3D emitter distribution, despite
reports of the 2D nature of the confinement by Munekata et al.
\cite{Munekata 1989}, and also in similar structures by Gonzalez
\cite{Gonzalez 2000}. Positive charge buildup in the GaSb is also
ignored in the analysis of Yu et al. \cite {Yu 1990} However, the
observation of intrinsic bistability in an
InAs/AlGaSb/InAs/AlGaSb/InAs double barrier structure \cite{Chow
1994} strongly suggests that both positive and negative charge
buildup must be included in any realistic analysis of conduction
in this family of devices.

In this work, we present vertical transport results on MOVPE grown
 DHETs with 'GaAs like' interfaces over a wider range of GaSb layer thickness
(0\,\AA\, - 1200\,\AA) than reported by Yu et al. \cite{Yu 1990}
The results from thicker GaSb layers are qualitatively similar to
those given in the literature, but here we report the first
observation of a dramatic drop in the conductance for DHETs with
GaSb thickness $<$ 60\,\AA. DHETs in this regime are shown to
behave dramatically differently to those with GaSb thickness $>$
60\,\AA. Hydrostatic pressures of up to 11 kbar are applied to
selected structures, allowing the continuous tuning of the band
overlap. This tuning of the relative electron and hole sub-band
alignments allows the electron-light-hole resonance to be pushed
below the region of band overlap and its role in the conduction to
be more carefully assessed than can be done by the GaSb thickness
dependence studies alone. An interband electroluminescence is also
observed with a threshold at the NDR valley bias. This provides
the first direct evidence of the nature of the background
conduction mechanism in these structures, previously of
unconfirmed origin.

We also present self-consistent calculations of the band profiles
to compare with our results. Our profiles appear to be
dramatically different to those found in the literature.
\cite{Ting 1992} We argue that this discrepancy has its origins in
the significant 2D hole concentration that accumulates in the
central GaSb layer. Analysis of the thickness and hydrostatic
pressure dependence of the conduction using this model provides an
adequate explanation for the mid infra-red emission beyond NDR,
and also gives strong evidence against the resonant
electron-light-hole tunnelling theory of Yu et al. \cite{Yu 1990}
Although the alignment of the light-hole sub-band is shown to be
important, it appears more likely that the conduction arises out
of an inelastic interaction between the electrons and heavy-holes,
more similar to that reported for SHETs with 'GaAs like'
interfaces. \cite{Khan-Cheema 1996}

\section{Experimental details}
n-InAs/p-GaSb/n-InAs nominally undoped DHETs were grown on
nominally undoped n-InAs substrates with GaSb layer thicknesses of
0-1200\,\AA, followed by a 3000\,{\AA} InAs cap. The interface
bonding was biased to be 'GaAs like' using a shutter sequence
similar to that described elsewhere \cite{Klipstein 1998}. The
layer thicknesses were calibrated by monitoring the Fabry-Perot
oscillations in the surface photoabsorption (spa) signal during
deposition \cite{Klipstein 1998}.

The wafers were processed into 150\,${\mu}$m mesa structures and
metalised with thermally evaporated, non annealed Cr/Au pads. 3-20
$\mu$sec pulsed I-V and electroluminescence were measured at
temperatures of 25-300 K. Pulsed I-V measurements were also
carried out under hydrostatic pressures up to 11 kbar using a
pressure cell as described elsewhere\cite{Khan-Cheema thesis}. All
DHETs showing NDR were highly sensitive to a small series
parasitic resistance which was always present due to contact
resistances of $\lesssim1\Omega$. This parasitic resistance could
be estimated fairly accurately and its effects removed by
calculating its two limiting values.  Too high a value gives rise
to a negative slope it the high current post NDR region while too
low a value leads to a peak bias lying above the valley.  Thus for
example the parasitic resistance for the $300$\,\AA DHET in
Fig.\ref{fig:IV dhets} was found to be in the range
$0.30\pm0.05\Omega$.

\section{Experimental Results}
\subsection{GaSb width dependence of I-V}
\begin{figure}[h]\centering \epsfxsize=85mm
\epsffile{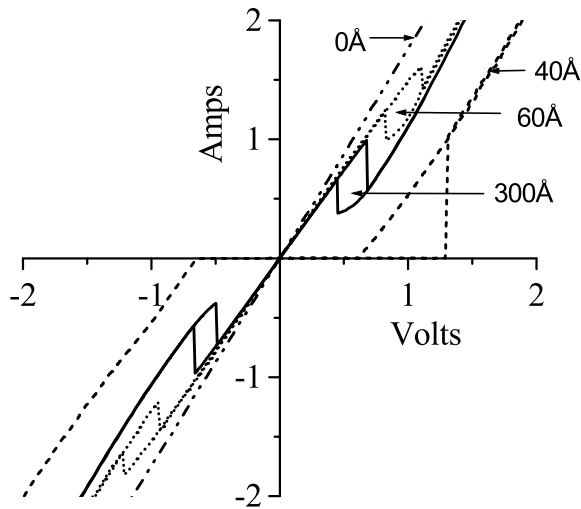} \vspace{0.2cm} \caption{I-V
characteristics of InAs/GaSb/InAs DHETs with different GaSb widths
at 77 K} \label{fig:IV dhets}
\end{figure}

The I-V results at 77 K as a function of GaSb thickness are
presented in Fig.\ref{fig:IV dhets}. The hysteresis in the NDR is
a result of the large current densities characteristic of these
structures making the effects of parasitic resistance significant
as already mentioned in section II. As a result, the voltage shown
does not represent the voltage across the DHET alone but includes
that across the parasitic resistance. However, the peak and valley
current densities are unaffected by the parasitic resistance.
These are found to scale with cross sectional mesa area as would
be expected for negligible surface leakage currents. The I-V
traces are only weakly dependent on GaSb thickness for GaSb

thickness above 60\,\AA, showing a slight decrease in peak and
valley currents for wider structures, similar to the results
reported by Yu et al. \cite{Yu 1990} For GaSb thickness below
60\,\AA, the I-V trace changes dramatically. This is most
remarkable for the 40\,\AA\, GaSb DHET example shown, where the
NDR is lost entirely at 77K and the current densities at low bias
are reduced by three orders of magnitude. Similar characteristics
were also observed for structures with GaSb thicknesses of 20, 30,
and 50\,\AA, some showing a marked hysteresis, and others none.
The hysteresis shown in Fig.\ref{fig:IV dhets}, which is also
observable under DC biasing, is not thought to be an effect of the
parasitic resistance as the device resistance is now much higher.
At the switching point the current density increases abruptly by
three orders of magnitude. The 0\,\AA\, control sample and the
10\,\AA\, GaSb DHET give an ohmic characteristic.

\subsection{Temperature dependence of I-V}
\begin{figure}[h]\centering \epsfxsize=85mm
\epsffile{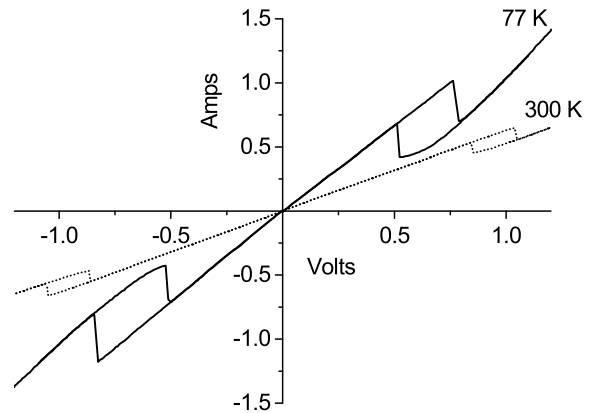} \vspace{0.2cm} \caption{I-V
characteristics of 150\,\AA\, GaSb DHET at 77 K and 300 K}
\label{fig:Temperature 150A}
\end{figure}

\begin{figure}[h]\centering \epsfxsize=85mm
\epsffile{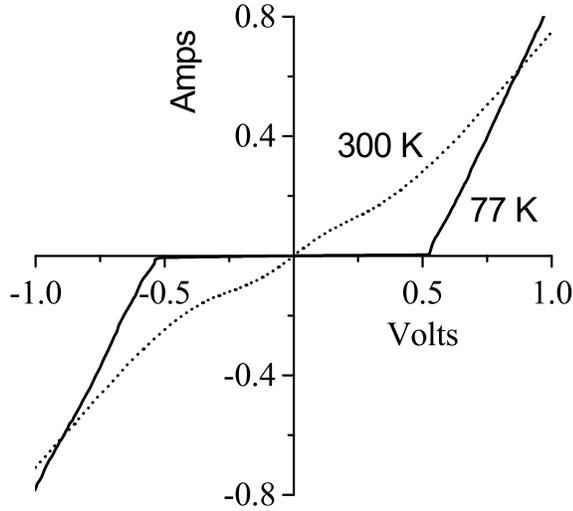} \vspace{0.2cm} \caption{I-V
characteristics of 40\,\AA\, GaSb DHET at 77 K and 300 K}
\label{fig:Temperature 40A}
\end{figure}

The temperature dependence for two DHETs with 150\,\AA\, and
40\,\AA\, of GaSb are shown in Fig.\ref{fig:Temperature 150A} and
in Fig.\ref{fig:Temperature 40A} respectively. The temperature
dependence for DHETs with GaSb thickness $>$ 60\,\AA\, is weak,
where the NDR is found to persist, as was the case for SHETs with
'GaAs like' interfaces \cite{Khan-Cheema 1996}. The shift along
the voltage axis is a result of changes in parasitic resistance
with temperature, confirmed by the 0\,\AA\, GaSb control sample
and consistent with reports in the literature. In contrast, the
temperature dependence for DHETs with GaSb thickness $<$ 60\,\AA\,
is much stronger. The dramatic suppression of the low bias
conductance in this regime is lifted at higher temperatures, with
a weak NDR feature returning around 200mV at 300 K.

\subsection{Hydrostatic pressure dependence of IV}
\begin{figure}[h]\centering \epsfxsize=85mm
\epsffile{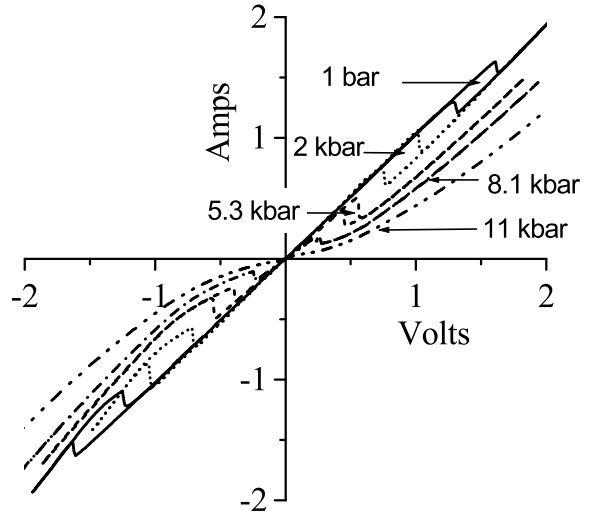} \vspace{0.2cm} \caption{IV
characteristics of 60\,\AA\, GaSb DHET at 77 K as a function of
hydrostatic pressure} \label{fig:IV pressure 60A}
\end{figure}

\begin{figure}[h]\centering \epsfxsize=85mm
\epsffile{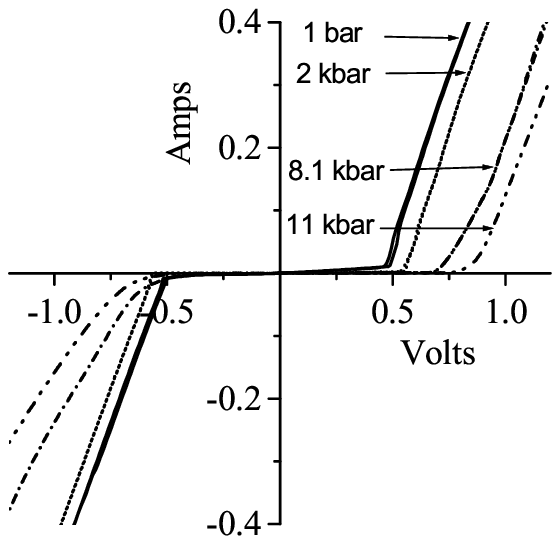} \vspace{0.2cm} \caption{IV
characteristics of 50\,\AA\, GaSb DHET  at 77 K as a function of
hydrostatic pressure} \label{fig:IV pressure 50A}
\end{figure}

The evolution of the I-V traces under hydrostatic pressure is
shown in both types of structure in Fig.\ref{fig:IV pressure 60A}
for the 60\,\AA\, GaSb DHET and in Fig.\ref{fig:IV pressure 50A}
for the 50\,\AA\, GaSb DHET. The NDR feature for the 60\,\AA\,
GaSb DHET moves continuously towards zero and is still discernible
at 11 kbar. For the thinner structures, the onset of conduction
increases from $\sim$500 mV to $\sim$750 mV between 1 bar and 11
kbar, after which the slope conductances are very similar. The
current axis is expanded in Fig.\ref{fig:IV pressure 50A zoom} to
reveal the strong suppression of the low bias conductance with
pressure for this structure.
\\

\begin{figure}[h]\centering \epsfxsize=85mm
\epsffile{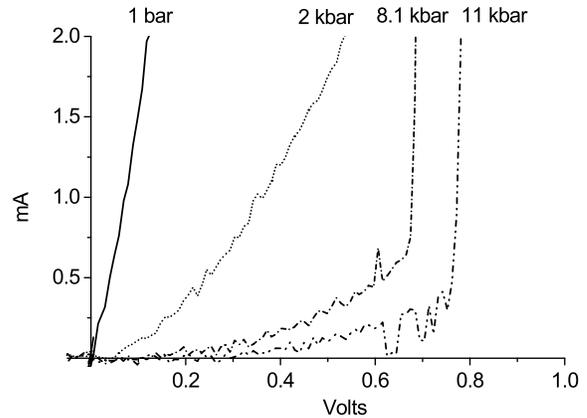} \vspace{0.2cm} \caption{IV
characteristics of 50\,\AA\, GaSb DHET  at 77 K as a function of
hydrostatic pressure expanded at low currents} \label{fig:IV
pressure 50A zoom}
\end{figure}

\subsection{Electroluminescence results}
\begin{figure}[h]\centering \epsfxsize=85mm
\epsffile{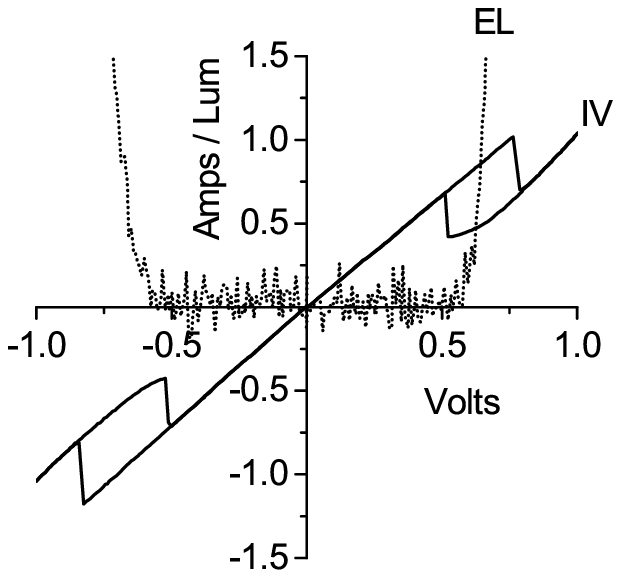} \vspace{0.2cm} \caption{I-V and
electroluminescence characteristics of a 150\,\AA\, GaSb DHET at
30 K} \label{fig:EL 150A}
\end{figure}

\begin{figure}[h]\centering \epsfxsize=85mm
\epsffile{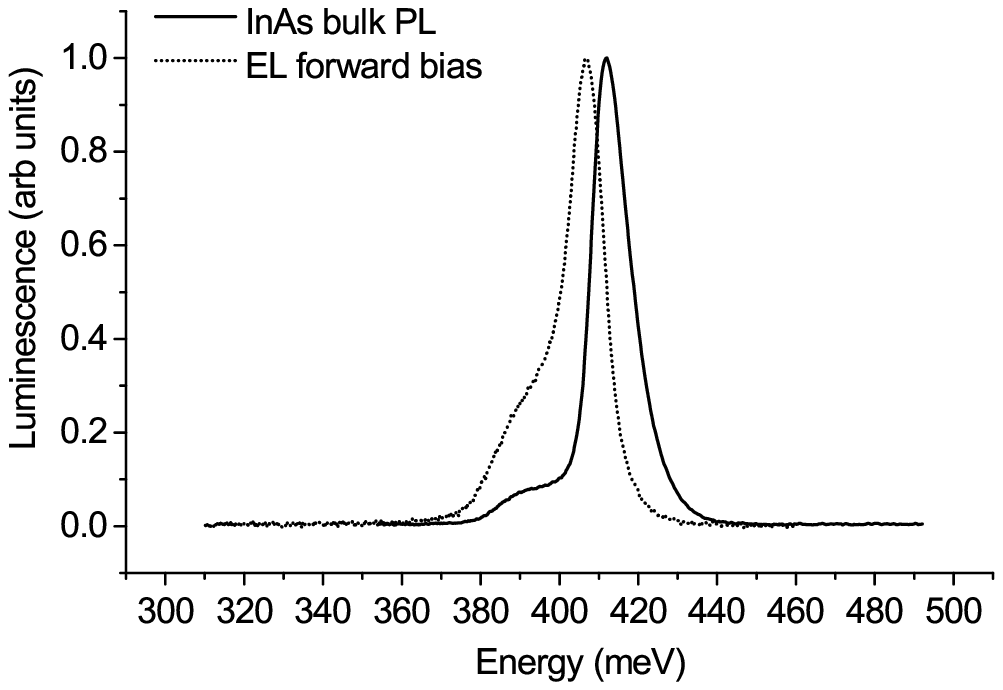} \vspace{0.2cm}
\caption{Photoluminescence and electroluminescence spectrum of
150\,\AA\, GaSb DHET at 30K} \label{fig:Lum spectrum}
\end{figure}

Mid IR electroluminescence is also detected with a threshold at
the NDR valley bias for all the structures showing NDR, as shown
for example by the 150\,\AA\, DHET in Fig.\ref{fig:EL 150A}. The
30 K electroluminescence spectrum corresponds to that of bulk InAs
photoluminescence, (peak at 410 meV or 3.02 ${\mu}$m), as shown in
Fig.\ref{fig:Lum spectrum}. The slight redshift in the
electroluminescence spectrum is attributed to Joule heating
effects. A similar emission is observed beyond 500mV for the
40\,\AA\, DHET.

\section{Band profile calculations}
Analysis of these results requires careful modelling of the band
profiles and sub-band energy levels both under equilibrium and non
equilibrium conditions. Self-consistent calculations on DHET
structures in the literature \cite{Ting 1992},\cite{Ting 1990}
ignore any confinement effects in the emitter, and use the 3D
Thomas-Fermi approximation to relate the Fermi level to the band
edges. Also, the band overlap appears to be independent of strain
and interface bonding, which we believe will have a dramatic
effect on the band profile calculations.

In our work, the emitter and collector are treated as 2D wells for
electrons, and the GaSb layer is treated as a 2D well for light
and heavy-holes, consistent with the true band bending expected
for a type II interface. As a first approximation, each layer in
this three layer system is assumed to be in local thermal
equilibrium, allowing a quasi-Fermi level to be defined in each.
Each layer is solved self-consistently using a transfer matrix
technique, under the assumption of total charge neutrality across
the whole structure. The band profile solved in this way for a
60\,\AA\, GaSb DHET at equilibrium is shown in
Fig.\ref{fig:Profile 60A}. The band offset is taken as 125meV for
the heavy-holes \cite{Khan-Cheema 1996}, \cite{Daly 1995a} and the
light-holes offset is lowered by $\sim$47meV as the GaSb is under
biaxial compression. Similar band profiles and experimental
evidence for the 2D nature of the electrons in the contact regions
has been provided in an analogous structure by Gonzalez et al.
\cite{Gonzalez 2000}.

\begin{figure}[h]\centering \epsfxsize=85mm
\epsffile{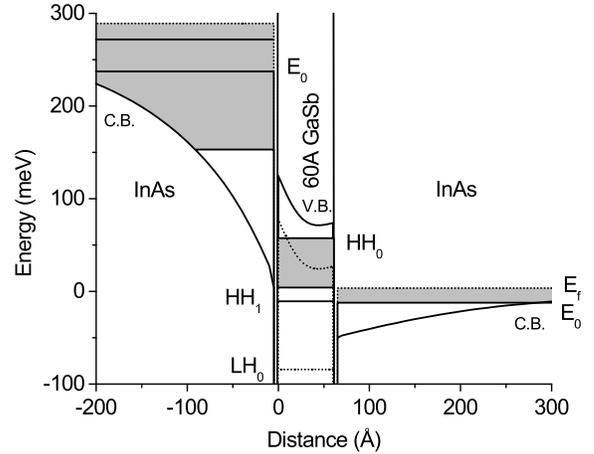} \vspace{0.2cm} \caption{Band
profile of 60\,\AA\, GaSb DHET at 285mV, close to the estimated
NDR peak bias across the structure} \label{fig:Profile 60A 285mV}
\end{figure}

Solving the band profile under non-equilibrium conditions requires
a much more complex analysis, as inelastic processes and
non-equilibrium Fermi distributions must be included in a
realistic model. In the analysis by Yu et al.\cite {Yu 1990}, all
inelastic processes are ignored and charge build-up in the GaSb
under biased conditions appears to be negligible, resulting in
potential profiles with a uniform electric field in the GaSb
region. In this work, the model used for the equilibrium band
profiles is adapted to give trial solutions under bias, which can
include both the confinement effects in the contact regions as
well as the effects of positive charge buildup in the central
layer. Self-consistent solutions correspond to cases where the
quasi-Fermi level of the holes in the GaSb lies in between those
of the contacts. The non-equilibrium band profile is found to be
very sensitive to the corresponding charge distributions. One
limiting solution gives band profiles very similar to those found
in the analysis of Yu et al. \cite{Yu 1990}, where the Fermi level
is broken at the collector interface and a depletion region is
formed at the collector contact. At the other extreme, the band
profile is shown in Fig.\ref{fig:Profile 60A 285mV} and resembles
that given by Lapushkin et al. \cite{Lapushkin 1997} in the
modelling of similar structures with AlSb barriers. In this case
the Fermi level is broken at the emitter contact, resulting in a
significant build-up of holes in the GaSb and an accumulation
region in the collector contact.

We point out that the first solution is highly unstable against
the rapid leakage of GaSb valence electrons into the InAs
conduction band at the $+$ve electrode and that the more likely
solution will resemble the band profile shown in
Fig.\ref{fig:Profile 60A 285mV}. This follows from estimates of
the light hole quasi bound state lifetime \cite{Ting 1992}, which
is predicted to be extremely short, of order a few tens of fsecs
for GaSb widths around 60-100\,\AA. This quasi-bound state
lifetime is related to the width of the electron-light-hole
transmission resonance, which is found to be much broader than for
typical quasi-bound states in more conventional double barrier
structures, due in the present case to the absence of barrier
layers. This short lifetime will promote the buildup of positive
charge in the central layer until the GaSb and collector contact
quasi Fermi levels are close to equilibrium, as is the case for
all the band profiles shown in this work.

Hole dispersions were calculated using a 4 band k.p model to give
an insight into the effects of valence band mixing. It turns out
that the large density of states offered by the flattening of the
HH$_{0}$ dispersion results in only the HH$_{0}$ sub-band being
occupied. Consequently, the light-hole resonance will not fall
between the contact Fermi levels in any of the cases studied. This
treatment therefore excludes any coherent tunnelling mechanism
through the electron-light-hole resonance, in contrast to the
theory of Yu et al. \cite{Yu 1990}

\section{Discussion}

We begin with the 3\,${\mu}$m luminescence observed in all samples
at biases $>$ 400mV. This is explained by the onset of Zener
tunnelling, which coincides with the NDR valley bias at ambient
pressure. Once the contact quasi-Fermi levels are separated in
energy by more than the InAs bandgap, valence electrons in the
emitter can tunnel into the conduction band of the $+$ve contact.
This explanation is consistent with the band profiles suggested in
this work, corresponding to that shown in Fig.\ref{fig:Zener
Schematic}. This observation shows clearly that this elastic Zener
tunnelling mechanism contributes significantly to the background
current in these structures, previously of unconfirmed origin.

\begin{figure}\centering \epsfxsize=85mm
\epsffile{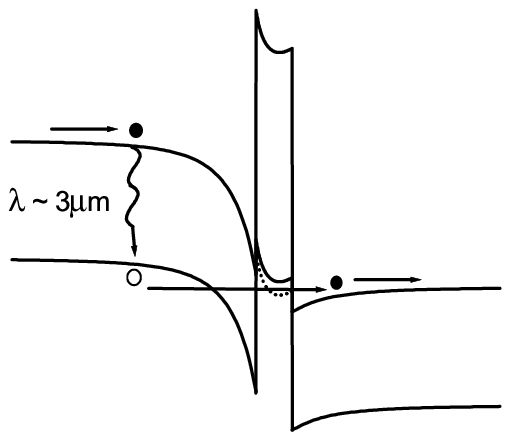} \vspace{0.2cm} \caption{Band
profile of 150\,\AA\, GaSb DHET at 500mV showing Zener tunnelling
mechanism and accompanying electroluminescence} \label{fig:Zener
Schematic}
\end{figure}

The peak NDR bias at 1 bar is now estimated and compared to
non-equilibrium band profiles to determine the nature of the peak
NDR conduction mechanism. The bias across the 60\,\AA\, DHET from
Fig.\ref{fig:IV pressure 60A} is estimated at $\sim$ 285 mV after
subtracting an estimated parasitic resistance of 0.8 $\Omega$
using the method described above. This is consistent with the
results from DHETs in the literature where parasitic resistances
are less dominant. The band profile given in Fig.\ref{fig:Profile
60A 285mV} shows that under these conditions the lowest electron
sub-band is far from being resonant with any of the hole
sub-bands. The electron and hole ground states are separated by
over $\sim$100 meV, suggesting that an inelastic process is
responsible for the NDR. This is in agreement with the analysis of
NDR in 'GaAs like' SHETs reported previously \cite{Khan-Cheema
1996}. Although the light-holes do not appear to be resonant with
the emitter electron distribution, their proximity will
nevertheless have the effect of increasing the penetration of the
electron wavefunction into the GaSb through electron-light-hole
mixing. Similarly, the penetration of the heavy-holes at finite
$k_{\parallel}$ into the InAs will be enhanced by the proximity of
the light-hole sub-band. The resulting electron-heavy-hole
wavefunction overlap will therefore be significant in this case
and allow an appreciable inelastic current to flow. Further
increase in the bias will result in a greater energy separation of
the electrons and holes and a stronger attenuation of the
wavefunction tails into the GaSb, at which point the region of NDR
is observed.

The hydrostatic pressure results for the 60\,\AA\, GaSb DHET are
consistent with this interpretation, and provide conclusive
evidence that the NDR does not arise out of resonant conduction
through the light-hole  sub-band. Hydrostatic pressure increases
the bulk band gaps and decreases the band overlap at $\sim$10
meV/kbar \cite{Daly 1995b} resulting in a lowering of the light
hole sub-band energy. At 5.3 kbar a clear NDR is observed despite
the energy of the light hole sub-band being unambiguously below
the region of band overlap, as shown in Fig.\ref{fig:Profile 60A
5kbar}.

\begin{figure}[h]\centering \epsfxsize=85mm
\epsffile{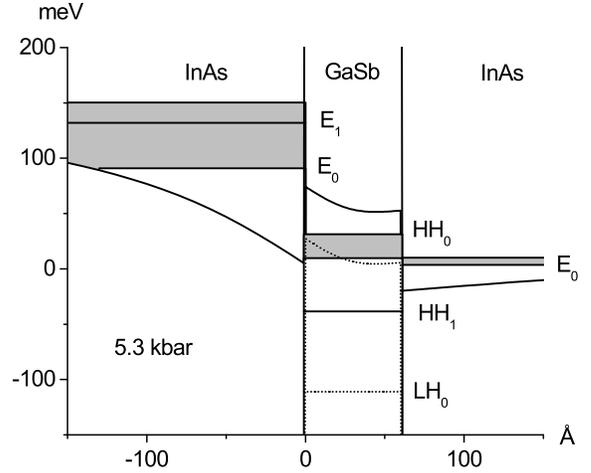} \vspace{0.2cm} \caption{Band
profile of 60\,\AA\, GaSb DHET at 141mV under 5.3 kbar, close to
the estimated NDR peak bias across the structure}
\label{fig:Profile 60A 5kbar}
\end{figure}

As the band overlap decreases with pressure, the charge transfer
across the interface also decreases, resulting in a change in
shape of the self-consistent potential and a lowering of the
confined electron sub-bands. This results in a reduced bias to
achieve the same electron-hole energy separation, hence the
decrease observed in the NDR position with increasing pressure
(Fig.\ref{fig:IV pressure 60A}) and the decrease in peak current.
The peak NDR position analysis of the band profiles for all of the
pressures studied shows that the electrons and heavy holes are
always uncrossed by $\sim$50-100 meV at the peak bias, again
consistent with an inelastic conduction mechanism.

We now turn to the dramatically reduced conduction in structures
with GaSb layers of 50\,\AA\, and less. The fact that hydrostatic
pressure does not induce similar behaviour in the 60\,\AA\, and
wider structures shows that the electron-hole level alignment and
overlap is much more strongly affected by confinement than by
pressure once the GaSb width is reduced below 60\,\AA. This is
indeed confirmed by the band profile calculations, where the
light-hole ground state is lowered by over 120meV as the GaSb
width changes from 80 to 40\,\AA\, whereas the electron sub-band
energy is hardly affected. This explains the dramatic change in
conduction observed, which we argue is a result of the
dramatically reduced electron-hole overlap as the light hole
energy is decreased. Application of hydrostatic pressure further
decreases the overlap as expected, and is reflected in the
reduction in current density in Fig.\ref{fig:IV pressure 50A
zoom}.

Moreover, the reappearance of a weak resonant feature at 1 bar at
300 K provides additional evidence for the inelastic nature of the
conduction in these structures. The increase in temperature gives
the opposite effect to pressure, with the band overlap increasing
by around 30meV between 77K and 300K \cite{Symons 1995} resulting
in a raising of the light hole sub-band relative to the emitter
electron distribution. Furthermore, the higher temperature will
allow holes to populate higher sub-bands with more light hole
character and greater penetration. The resulting enhanced
electron-hole overlap then gives rise to an increase in the
conductivity, and even a weak NDR, as observed.

Finally, we consider the intrinsic bistability shown in many of
the samples studied, where the conduction is found in some cases
to abruptly switch by over three orders of magnitude. This
strongly suggests that there are two different steady state charge
distributions at a given bias, and we propose that these
correspond to there being different amounts of positive charge
accumulation in the GaSb, in some ways similar to that proposed by
Chow et al. \cite{Chow 1994} Although a significant positive
charge buildup is very likely in the thicker structures, it is
feasible that this is not always the case for thinner structures.
The efficient leakage of electrons from the GaSb valence band into
the positive contact could indeed become more difficult for
thinner structures as the light hole sub-band is pushed to lower
energies. When the GaSb barrier becomes even thinner, direct
electron tunnelling may begin to occur.

The conduction mechanisms must be dramatically different in the
two bistable configurations to give such radically different
current densities, clearly demonstrating the sensitivity of these
structures to the band profile models used.

\section{Conclusions}
We have presented a detailed investigation into the nature of the
conduction in InAs/GaSb/InAs DHETs. A dramatic drop in the
conductance, a loss in the NDR and marked intrinsic hysteresis is
observed for DHETs with GaSb thickness $<$60\,\AA. The thickness
and hydrostatic pressure dependence of the conduction are
interpreted with the help of self-consistent band profiles, where
the inclusion of a significant positive charge buildup in the GaSb
is found to give solutions quite different to those found in the
literature. Resonant electron-light hole tunnelling is shown not
to be the dominant transport mechanism, although the position of
the light-hole sub-band is believed to be important in determining
the degree of electron-hole overlap necessary for conduction.
Inelastic conduction involving electrons and heavy-holes is most
likely to be the dominant process at the peak NDR bias, similar to
the conduction mechanism proposed for the InAs/GaSb SHETs with
'GaAs like' interface bonding \cite{Khan-Cheema 1996}. The
hysteresis observed in many of the DHETs with thinner GaSb layers
demonstrate that the conduction is very sensitive to the steady
state charge distribution. An interband luminescence observed
beyond NDR is explained by Zener tunnelling, confirming an
important contribution to the background conduction mechanism.

\section{Acknowledgements}
We are grateful to the EPSRC (UK) for the support of this work.

\end{document}